\documentclass[aps, reprint, twocolumn, superscriptaddress, citeautoscript]{revtex4-1}

\usepackage{graphicx}

\usepackage{amssymb}
\usepackage{amsmath}
\usepackage{amsfonts}
\usepackage{mathrsfs}

\usepackage{bm}
\usepackage{dcolumn}
\usepackage{color}
\usepackage[colorlinks=true,citecolor=blue]{hyperref}
\hypersetup{colorlinks=true,citecolor=blue,linkcolor=red, urlcolor=blue}

\usepackage{float}

\usepackage{color}
\usepackage{ulem}
\definecolor{purple}{rgb}{0.8,0,0.6}
\definecolor{orange}{rgb}{1,0.55,0}

\newcommand{\beqn}{\begin{eqnarray}}
	\newcommand{\eeqn}{\end{eqnarray}}

\newcommand{\comment}[1]{}

\begin{document}
	
	\preprint{AIP/123-QED}
	
	\title[]{Interplay of Electron Trapping by Defect Midgap State and Quantum Confinement to Optimize Hot Carrier Effect in a Nanowire Structure}
	
	\author{Imam Makhfudz}
	\affiliation{IM2NP, UMR CNRS 7334, Aix-Marseille Universit\'{e}, 13013 Marseille, France}
	\author{Hamidreza Esmaielpour}
	\affiliation{Walter Schottky Institut, Technische Universität München, Am Coulombwall 4, D-85748 Garching, Germany}
	\author{Yaser Hajati}
	\affiliation{Department of Physics, Faculty of Science, Shahid Chamran University of Ahvaz, 6135743135 Ahvaz, Iran}
	\author{Gregor Koblmüller}
	\affiliation{Walter Schottky Institut, Technische Universität München, Am Coulombwall 4, D-85748 Garching, Germany}
	\author{Nicolas Cavassilas}
	\affiliation{IM2NP, UMR CNRS 7334, Aix-Marseille Universit\'{e}, 13013 Marseille, France}

	\date{\today}
	
	\begin{abstract}
		Hot carrier effect, a phenomenon where charge carriers generated by photon absorption remain energetic by not losing much energy, has been one of the leading strategies in increasing solar cell efficiency. Nanostructuring offers an effective approach to enhance hot carrier effect via the spatial confinement, as occurring in a nanowire structure. The recent experimental study by Esmaielpour et al. [ACS Applied Nano Materials 7, 2817 (2024)] reveals a fascinating non-monotonic dependence of the hot carrier effect in nanowire array on the diameter of the nanowire, contrary to what might be expected from quantum confinement alone. We show that this non-monotonic behavior can be explained by a simple model for electron energy loss that involves two principal mechanisms. First, electron-phonon scattering, that increases with nanowire diameter, leading to hot carrier effect that decreases with increasing diameter. Second, electron capture by a defect level within band gap, that is, a midgap state, that decreases with nanowire diameter, leading to hot carrier effect that increases with increasing diameter. The two mechanisms balance at a certain diameter corresponding to optimal hot carrier effect. Our result offers a guideline to optimize hot carrier effect in nanowire solar cells and ultimately their efficiency by adjusting the dimensions and micro-structural properties of nanowires.
	\end{abstract}
	
	\maketitle
	
	\textit{Introduction.\textemdash}One of the main objectives of third generation photovoltaic solar cells is to design devices with efficiencies above the theoretical limit in single-junction solar cells (33$\%$), known as Shockley-Queisser limit \cite{ShockleyQueisser}. In photovoltaic devices, various physical processes contribute to the reduction of the solar cell efficiency. One of them is electron energy loss due to scattering by lattice vibrations, motivating proposals to reduce it, such as engineering of phonon density of states \cite{MahyarAPL} and confinement of the electron in a quantum well \cite{NC}\cite{MakhfudzJoPD} or superlattice \cite{MakhfudzPhysRevApplied}. The goal is to keep the electrons remain energetic by not losing much of their kinetic energy, thus staying ``hot'', up to the point that they get extracted out of the device by external circuit. In this case, the electrons are considered hot because their effective temperature becomes significantly higher than the ambient (lattice) temperature, can be by tens to hundreds of Kelvin. First suggested by Ross and Nozik \cite{RossNozik}, this phenomenon called hot carrier effect \cite{JShah} has now become one of the leading strategies to enhance solar cell efficiency \cite{Conibeer1,Conibeer2,BrisGuillemoles, DemoHCSC}.
	
	Among various designs for hot carrier absorbers, nanostructures have shown promising results in improving the effect of hot carriers under lower solar concentrations compared to bulk structures. This intriguing effect is attributed to the spatial confinement of charged particles in nanostructures and the adjustment of material properties, which can slow down the rates of hot carrier thermalization. However, the presence of non-idealities in the crystal structures, like defects, can influence the properties of hot carriers, resulting higher rates of thermalization. Hereby, we elucidate the intertwining roles of electron-phonon scattering and defect-induced scattering of electron in controlling the carrier thermalization and hence the hot carrier effect in  semiconductor nanowire heterostructure. In particular, we demonstrate that the two mechanisms lead to opposite dependencies of electron energy loss rate on nanowire diameter, which should drive a strong dimensional and structural dependencies of hot carrier effect in this heterostructure, providing a solid explanation to a recent intriguing observation of non-monotonic hot carrier effect dependence on the diameter of the  nanowires \cite{MahyarACSnanoenergymaterials}.

	\textit{Electron-Phonon Scattering Mechanism.\textemdash}Electron-phonon scattering is one of the main channels of hot carrier thermalization in semiconductors. In polar semiconductor materials of our interest here, electrons lose energy by emission of longitudinal optical (LO) phonons, providing the principal way by which the electrons thermalize from higher energy states to the bottom of conduction band. 
	\begin{figure}
		\includegraphics[angle=0,origin=c, scale=0.8]{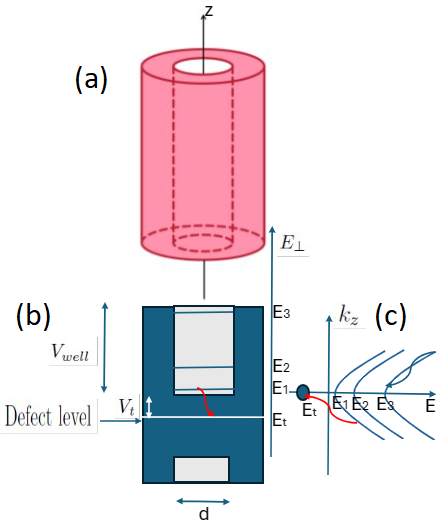}
		\caption{
			(a)The geometry of the nanowire with the core (inner white cylinder of radius $R$) enclosed by the shell. (b)The energy level diagram with  a defect level within the band gap (dark blue region) with discrete confined electron transverse energy levels $E_{\perp}\equiv \varepsilon_{l,n}=E_1,  E_2, E_3, \cdots$ (Eq.(\ref{EigenvaluesNW})) within the potential well (grey region) in the conduction band and (c)the corresponding total energy $E$ with its parabolic dispersion along $k_z$. The blue and red curly arrows respectively represent electron energy losses by phonon emission and trapping by defect level. }\label{fig:Fig1}
	\end{figure}
	
	We will compute the electron-phonon scattering rate using a simple model of electron-phonon scattering within kinetic theory. Employing standard Fr\"{o}hlich model of electron-phonon interaction \cite{Mahan}, the electron-phonon scattering rate is given by \cite{MakhfudzJoPD}
	\begin{equation}\label{RateofPhononNumberChange}
		\frac{d N_{{\bf q}_\perp,q _z}}{dt} = \frac{2 \pi}{\hbar} |M_{{\bf q}_\perp,q _z} |^2 A_{\mathbf{q}_{\perp},q_z}
	\end{equation}
	where $\hbar$ is the reduced Planck constant and the electron-phonon coupling function $M_{\mathbf{q}_\perp,q_z}$ is given by
	\begin{equation}	M_{\mathbf{q}_\perp,q_z}=\sqrt{\frac{\hbar\omega_{LO} e^2}{8\pi q^2 V}\left(\frac{1}{\varepsilon_{\infty}}-\frac{1}{\varepsilon_0}\right)}
	\end{equation}
	with $q=\sqrt{q^2_\perp+q^2_z}$ the wave vector of the LO phonon, while $\omega_{LO},e,V,\varepsilon_{\infty},\varepsilon_0$ are respectively the LO phonon frequency, elementary charge, phonon spatial volume, infinite frequency and static dielectric constants. For a nanowire of length $L$ and circular cross sectional area $S=\pi R^2$ as illustrated in Fig. \ref{fig:Fig1}, then $V=SL$. The nanowire is long enough that the electrons are effectively free particles along the nanowire axis (taken to be along the $z$ axis) with plane wave function. On the other hand, the electrons are confined in the transverse cross sectional $xy$ plane of the nanowire, and their energies may be quantized in this plane.
	
	The kinetic function $A_{\mathbf{q}_{\perp},q_z}$ is given by
	\[
	A_{\mathbf{q}_{\perp},q_z}=2\sum_{k_z, k'_z}\sum_{\mathbf{k}_{\perp},\mathbf{k}'_{\perp}}I^2_{\mathbf{k},\mathbf{k}'}((N_{\mathbf{q}}(T_L)+1)f_{\mathbf{k}_{\perp},k_z}(1-f_{\mathbf{k}'_{\perp},k_z-q_z})
	\]
	\begin{equation}\label{TheSum}
		-N_{\mathbf{q}}(T_L)f_{\mathbf{k}'_{\perp},k_z-q_z}(1-f_{\mathbf{k}_{\perp},k_z}))
		\delta(\Delta E_{\mathbf{k}(\mathbf{n})\mathbf{k}'(\mathbf{n})',\mathbf{q}}-\hbar\omega_{\mathbf{q}_{\perp},q_z})
	\end{equation}
	where $\Delta E_{\mathbf{k}(\mathbf{n});\mathbf{k}'(\mathbf{n}'),\mathbf{q}}=E_\mathbf{n}(k_z)-E_{\mathbf{n}'}(k_z-q_z)$ the energy of transition from level $\mathbf{n}$ to $\mathbf{n}'$ by emission of an LO phonon of energy $\hbar\omega_{\mathbf{q}_{\perp},q_z}=\hbar\omega_{LO}$ (assumed to be constant independent of its wave vector $\mathbf{q}$) as illustrated in Fig.\ref{fig:Fig1},  ${\bf k}(\mathbf{n}) = ({\bf k}_\perp(\mathbf{n}), k_z)$ for which $\mathbf{n}$ is a set of two quantum numbers characterizing the confined states of electron in the cross sectional plane of the nanowire, $N_{\mathbf{q}}(T_L)$ is the number of LO phonons of wave vector $\mathbf{q}=(\mathbf{q}_{\perp},q_z)$ given by the Bose-Einstein distribution $N_{\mathbf{q}}(T_L)=1/(\exp(\hbar \omega_{LO}/(k_BT_L))-1)$ at lattice temperature $T_L$, and $f_{\zeta}=(\exp((E_{\zeta}-\mu_c)/(k_BT_c))+1)^{-1}$ where  $k_B$ is the Boltzmann constant; the Fermi-Dirac distribution of electron of wave vector $\zeta$ having energy $E_{\zeta}$ at temperature $T_c$ and chemical potential $\mu_c$. The sums in equation (\ref{TheSum}) involves sums over $k_z, k'_z$, the electron momenta along the axis $z$ of the nanowire respectively before and after scattering, corresponding to that of a free particle electron with plane wave function. Momentum conservation along this direction eventually imposes $k'_z=k_z-q_z$, as appearing in the subscript of the Fermi-Dirac functions. On the other hand,  the sums over transverse wave vectors involve quantized wave vectors $\mathbf{k}_{\perp}(\mathbf{n}), \mathbf{k}_{\perp}(\mathbf{n}')$ due to electron confinement in this plane. The $I^2_{\mathbf{k},\mathbf{k}-\mathbf{q}}=|G(\mathbf{k}(\mathbf{n})\mathbf{k}'(\mathbf{n})\mathbf{q})|^2$ where $G(\mathbf{k}(\mathbf{n})\mathbf{k}'(\mathbf{n})'\mathbf{q})$ is the ``form factor" obtained from the overlap of the electron wave functions before and after its scattering by an LO phonon
	\begin{equation}\label{formfactor}
		G(\mathbf{k}(\mathbf{n})\mathbf{k}'(\mathbf{n})'\mathbf{q})=\delta(k'_z-k_z+q_z)\int d^2r \phi^*_{\mathbf{n}}(\mathbf{r})e^{i\mathbf{q}\cdot\mathbf{r}}\phi_{\mathbf{n}'}(\mathbf{r})
	\end{equation}
	where $\phi_{\mathbf{n}}(\mathbf{r})$ is the confined electron wave function in the plane of the circular cross section of the nanowire. The Dirac delta function reflects momentum conservation along the $z$ axis and the associated translational invariance along the $z$ axis of the nanowire.
	
	We will consider the simplest form of Hamiltonian within approximation of effective mass $m_c$ to model the electrons in the nanowire: electron confined in a cylindrical potential, first approximated with infinitely high potential wall. The Schrodinger equation is given by
		\begin{equation}
			\left[ -\frac{\hbar^2}{2m_c}\nabla^2+V(r,\phi)\right]\psi(r,\phi,z)=E\psi(r,\phi,z)
		\end{equation}
	with standard solution, in cylindrical coordinates \cite{SemiconBook}
	\begin{equation} \label{electroneigenfunctioninnanowire}
		\psi(r,\phi,z)\equiv \phi_{\mathbf{n}}(\mathbf{r})=J_l\left(\frac{\alpha_{l,n}r}{R}\right)e^{il\phi}e^{ik_zz},
	\end{equation}
	\begin{equation}\label{EigenvaluesNW}
		E \equiv E_\mathbf{n}(k_z)=\frac{\hbar^2k^2_z}{2m_c}+\frac{\hbar^2\alpha^2_{l,n}}{2m_cR^2}=\frac{\hbar^2k^2_z}{2m_c}+\varepsilon_{l,n}
	\end{equation}
	where $l=0, \pm 1, \pm 2, \cdots$ to be referred as ``orbital'' quantum number and  $n=1,2,3,\cdots$ the node quantum number. In this context of circular nanowire, therefore $\mathbf{n}=(l, n)$. The eigenfunctions and eigenvalues above are based on modeling the nanowire as an infinite potential well $V(r,\phi)=0$ for $r<R$ while $(r,\phi)=\infty$ for $r>R$, corresponding to hard wall boundary condition, which is justified when the actual depth of the wall is far larger than the energy level spacing. The $J_l(\cdots)$ is Bessel function of the first kind of the $l^{th}$ order, $\alpha_{l,n}$ is its $n^{th}$ zero; i.e. $J_l(\alpha_{l,n})=0$. Equation (\ref{EigenvaluesNW}) also implies that
	
	\begin{equation}
		\mathbf{k}^2_{\perp}(\mathbf{n})=\frac{\alpha^2_{l,n}}{R^2}
	\end{equation}
	representing the quantization of electron wave vector in the transverse (cross section) plane of the nanowire.
	
	The density of states (per unit energy per unit volume) $N(E)$ corresponding to the solutions (\ref{electroneigenfunctioninnanowire}-\ref{EigenvaluesNW}) is given by \cite{SemiconBook}
	\begin{equation}\label{1DelectronDOS}
		N(E)=\frac{1}{\pi R^2}\sum^{\infty}_{l=-\infty}\sum^{\infty}_{n=1}\frac{1}{\pi \hbar}\sqrt{\frac{2m_c}{E-\varepsilon_{l,n}}}\Theta(E-\varepsilon_{l,n})
	\end{equation}
	where $\Theta(\cdots)$ is the Heaviside step function; $\Theta(x)=1$ for $x\geq 0$ and zero otherwise. 
	
	Evaluating Eq.(\ref{RateofPhononNumberChange}) using equations (\ref{TheSum}-\ref{formfactor}) and following the standard derivation of scattering rate within kinetic theory \cite{MakhfudzJoPD} but now adapted to a cylindrical geometry, it can be shown that \cite{SupplementaryMaterials} 
	\[
	\frac{d N_{{\bf q}_\perp,q _z}}{dt} =\frac{N_{\mathbf{q}}(T_c)-N_{\mathbf{q}}(T_L)}{\tau^{c-LO}_{\mathbf{q}}}
	\]
	derived employing an identity \cite{WurfelMain} involving electron Fermi-Dirac distribution and the Bose-Einstein distribution $N_{\mathbf{q}}(T_c)=1/(\exp(\hbar \omega_{LO}/(k_BT_c))-1)$\cite{SupplementaryMaterials},  giving an electron-phonon scattering rate
	\[
	\frac{1}{\tau^{c-LO}_{\mathbf{q}}}=
	\]
	\begin{equation}
		\frac{2Lm_c}{\hbar^3q_z}|M_{\bf{q}_{\perp},q_z}|^2\sum_{\mathbf{k}_{\perp},\mathbf{k}'_{\perp}}I^2_{\mathbf{k}_{\perp},k^0_z;\mathbf{k}'_{\perp},k^0_z-q_z}\left(f_{\mathbf{k}'_{\perp},k^0_z-q_z}-f_{\mathbf{k}_{\perp},k^0_z}\right)
	\end{equation}
	where 
	\begin{equation}
		k^0_z=\frac{1}{2q_z}\left(\mathbf{k}'^2_{\perp}+q^2_z-\mathbf{k}^2_{\perp}\right)+\frac{m_c}{\hbar^2q_z}\hbar\omega_{LO}
	\end{equation}
and the sums over $\mathbf{k}_{\perp},\mathbf{k}'_{\perp}$ in are eventually subjected to the constraint \begin{equation}\label{constraintonenergy}
	\varepsilon_{l,n}, \varepsilon_{l',n'}\leq V_{well}
\end{equation}
where $V_{well}$ is the actual confining potential energy of the nanowire (rather than infinity as we assumed in the first step of the approximation) to take into account of the fact that there are only finite number of confined electron states in real systems. The hard wall boundary condition and the energy level truncation scheme above are justified since the condition $V_{well}\gg \Delta E$, where $\Delta E = \hbar^2/(2m_cR^2)$ is energy level spacing, is satisfied in the calculations presented in this work, using parameters that describe realistic nanowire system. 

The thermalization power per unit cross sectional area, describing the rate of energy loss of the electrons from emitting LO phonons within the nanowire, is given by 
\begin{equation}\label{thermalpower3dbulk}
	P^{}_{\mathrm{th}}=L\int \frac{d^3q}{(2\pi)^3} \hbar\omega_{LO} \frac{N_{\mathbf{q}}(T_c)-N_{\mathbf{q}}(T_L)}{\tau^{c-LO}_{\mathbf{q}}}
\end{equation}
with the dimension of energy time $^{-1}$length$^{-2}$.

\textit{Electron-Defect Scattering: Shockley-Read-Hall Theory.\textemdash}
In nanowire with nanometer length scale, defect is a common problem. Defect may come from structural imperfection intrinsic from the material itself such as vacancy, but it may also come from the presence of external atom commonly called impurity. In both cases, we will refer to them as defect. We will in particular focus on the simplest  but yet most consequential type of defect; point defect, as it leads to midgap state acting as scattering center
\cite{WKohn}\cite{LannooBook1} acting as a trapper of electron as illustrated in Fig.\ref{fig:Fig1}, while other types of defects such as planar defect are known to be less influential \cite{NanoLettersPlanarDefect}. 

In Shockley-Read-Hall (SRH) statistical theory of non-radiative recombination of electrons and holes \cite{ShockleyRead}\cite{Hall}, the net electron capture rate is computed from detailed balance between the electron capture by the defect level and the electron emission back into the conduction band. While the original SRH theory was formulated for bulk system, we extend the theory to quantum confined geometry with discrete energy levels for the electron \cite{NoteDefect}. Writing the density of states in Eq.(\ref{1DelectronDOS}) as 
\begin{equation}
	N(E)=\sum^{\infty}_{l=-\infty}N_l(E),
\end{equation}
the net electron capture rate is then given by
\begin{equation}\label{electroncaptureRate}
	R_{ce}=\left[1-e^{\beta(F_t-F_n)}\right]f_{pt}N_t\int^\infty_{\epsilon_c}dE\sum^{\infty}_{l=-\infty}f(E)N_l(E)c_l(E)
\end{equation}
where `ce' is for `capture of electron',    which sums up the contributions of all orbital channels $l$, where $\beta=1/(k_BT_c)$, $F_n (\equiv \mu_c), F_t$ are respectively the quasi-Fermi levels corresponding to the electron occupation of the conduction band and trap state respectively, $f_{pt}=1-f_t$ where $f_t=(\exp((E_t-F_t)/(k_BT_c))+1)^{-1}$ is the Fermi-Dirac distribution function of the electron at the defect energy $E_t$ with Fermi energy $F_t$ where the subscript $t$ is for the `trap', $N_t$ is the defect volume density \cite{NoteDefectDensity}, $f(E)$ is the Fermi-Dirac probability distribution at general energy $E$ within the conduction band, $N_l(E)$ is the electron density of states from the orbital channel $l$, while $c_l(E)$ is the corresponding average probability of electron capture in that orbital channel per unit time. The latter is given by \cite{ShockleyRead}
\begin{equation}\label{electroncaptureprobability}
	c_l(E)=\langle v(E) A_l(E)\rangle
\end{equation}
where $v(E)$ is the velocity of the electron at energy $E$ while $A_l(E)$ is the scattering cross section, which we have defined for each orbital $l$, at energy $E$ and $\langle \dots \rangle$ indicates averaging with respect to appropriate quantities.  

An electron of energy $E$ in the conduction band falling into a trap level of energy $E_t$ loses energy $E-E_t$. Conversely, an electron escaping from the trap level and jump to the conduction band at energy $E$ acquires energy $E-E_t$. As such, the net electron energy loss rate (per unit area) due to the trap level in a circular nanowire of radius $R$ and length $L$ is given by
\begin{equation}\label{electronenergylossRate}
	P_{ce}=R_{ce}\frac{\int^\infty_{\epsilon_c}dE\sum^{\infty}_{l=-\infty}(E-E_t)f(E)N_l(E)c_l(E)}{\int^\infty_{\epsilon_c}dE\sum^{\infty}_{l=-\infty}f(E)N_l(E)c_l(E)}
\end{equation}
which is a positive definite quantity for $F_t<F_n$. This energy loss rate is the direct analog of the thermalization power given in Eq.(\ref{thermalpower3dbulk}).

\begin{table}
	\begin{tabular}{l|c|l}
		\hline\hline
		Quantity & Symbol & Value \\
		\hline
		LO phonon energy & $\hbar\omega_{LO}$ & $36$ meV \\
		Electron effective temperature & $T_c$ & $90$ K \\
		Lattice temperature & $T_L$ & $10$ K \\
		Conduction electron Fermi energy & $F_n \equiv  \mu_c$ & $0.0$ eV \\
		Defect electron Fermi energy & $F_t$ & $-0.25$ eV \\
		Defect level energy & $E_t$ & $-0.25$ eV\\
		Defect potential energy & $V_t$ &  0.25 eV\\
		Defect potential radius & $r_t$ & 0.35 nm\\
		Reference point defect density & $N_0$ & 3.7$\times 10^{-5}$/nm$^3$\\
		Reference nanowire diameter & $d_0$ & 200 nm\\
		Well depth for electrons & $V_{well}$ & 0.525 eV\\
		Electron effective mass (InGaAs) & $m_c$ & 0.041 $m_0$ \\
		Infinite-frequency susceptibility & $K_{\infty}$ & 12.9\\
		Static susceptibility & $K_s$ & 10.9\\
		\hline
	\end{tabular}
	\caption{\label{tbl:parameters} Material parameters used to compute the electron-LO phonon scattering rate and thermalization power in InGaAs/InAlAs core-shell nanowire. The defect density is taken to be $N_t=N_0d^2_0/d^2$ \cite{NoteDefectDensity}.}
\end{table} 

\begin{figure}
	\includegraphics[angle=0,origin=c, scale=0.6]{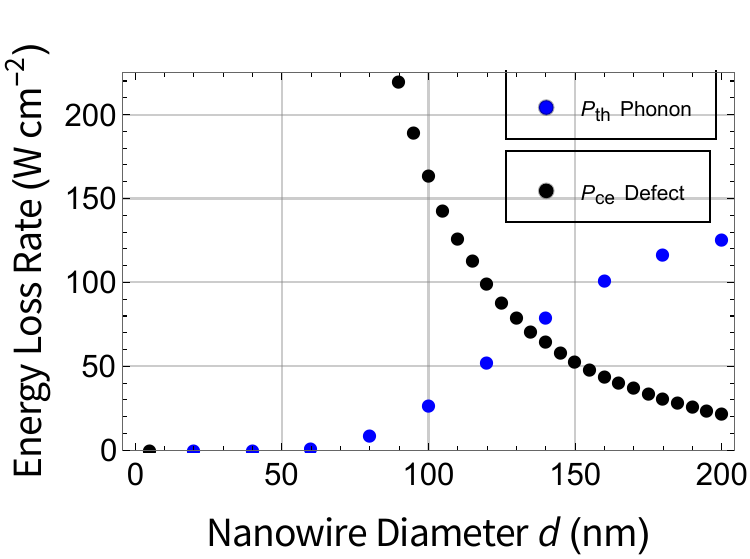}
	\caption{
		The theoretical dependence (black circles)  of the defect-induced net electron energy loss rate $P_{ce}$ on the nanowire diameter of a 1000 nm-long InGaAs/InAlAs core-shell nanowire, computed with parameters in Table I. Superposed (blue circles) is the electron-phonon scattering-induced thermalization power $P_{th}$.}\label{fig:EnergyLossRates}
\end{figure}

The point defect potential is modeled as a short-ranged spherical well of depth $V_t=-E_t$ and radius $r_t$. This model is appropriate to describe point defect that gives rise to `deep level' \cite{Keldysh}, also the one applicable to SRH theory \cite{ShockleyRead}. The differential cross section is then computed using partial wave technique \cite{LandauLifshitzQM} with details given in \cite{SupplementaryMaterials} within Born approximation and with spatial constraint $r_t\ll R$, giving
	\begin{equation}\label{ElectronCaptureProbability}
		c(E)=\frac{\pi^3\hbar^2V^2_t}{4p^4_t}\sqrt{\frac{2m_c}{E}}\sum^{\infty}_{l'=1}(2l'+1)^2G^2_{l'}(E)
	\end{equation} 
	for all $l$, where $p_t=\hbar/r_t$ and
	\begin{equation}
		G_{l'}(E)=J^2_{l'+\frac{1}{2}}(kr_t)-J_{l'-\frac{1}{2}}(kr_t)J_{l'+\frac{3}{2}}(kr_t)
	\end{equation}
	with
	\begin{equation}
		k=\frac{1}{\hbar}\sqrt{2m_c E}.
	\end{equation} 
	The approximation that leads to Eq.(\ref{ElectronCaptureProbability}) is also justified because the number of confined states is already large enough within the range of diameter ($\sim$ 50-200nm) that will be of our interest, rendering the quantization (in terms of $l,n$) less visible. 
	It can be verified that  $c(E)$ decreases with energy $E$, which indicates that the higher the electron energy is in the conduction band, the less likely it will be captured by the defect level. 
	
	Employing the result for the electron capture probability from Eq.(\ref{ElectronCaptureProbability}) into the expression for the net electron capture rate in Eq.(\ref{electronenergylossRate}), we obtain 
	\begin{equation}\label{electroncaptureRateFinal}
		P_{ce}=\left[1-\exp(\beta(F_t-F_n))\right]f_{pt}N_t I_{nE}
	\end{equation}
	where
		\[
		I_{nE}=\int^\infty_{\epsilon_c}dE\sum^{\infty}_{l=-\infty}(E-E_t)f(E)N_l(E)c(E)=\frac{m_c\pi\hbar V^2_t}{2p^4_tR^2}\
		\]
		\begin{equation}\label{IntegralFactor}
			\times\sum^{\infty}_{l=-\infty, l',n=1}(2l'+1)^2\int^{\infty}_{\varepsilon_{l,n}} dE\frac{(E-E_t)f(E)G^2_{l'}(E)}{\sqrt{E(E-\varepsilon_{l,n})}}      
		\end{equation}
	where the sums over $l,n$ are subject to the constraint Eq.(\ref{constraintonenergy}) and we have used the bottom of the bulk conduction band as reference energy $\epsilon_c=0$ and taken the density of states of electron in  Eq.(\ref{1DelectronDOS}). 
	
	\textit{Competing Roles of the Two Mechanisms.\textemdash}The resulting defect-induced electron energy loss rate $P_{ce}$ is displayed in Fig. \ref{fig:EnergyLossRates}, computed using parameters listed in Table I. The numerical values of intrinsic parameters of InGaAs are standard \cite{NoteParameters} and gives an electron density of the same order of magnitude as photogenerated electron density estimated from experiment \cite{MahyarACSnanoenergymaterials} ($\simeq 10^{22}$/m$^3$), whereas those characterizing the defect are estimated based on the results available in the literature for the energy parameters \cite{Komsa2010,Komsa2012a,Komsa2012b}, defect density from \cite{MahyarACSnanoenergymaterials}, while potential range $r_t$ is supposed to be smaller than typical lattice spacing $\sim 5.8\AA$, but this is not a necessary condition. As comparison, the electron-phonon $P_{th}$ is superimposed. While the number of confined electron states increases with the nanowire diameter, the black circles in Fig. \ref{fig:EnergyLossRates} however shows that the electron energy loss from trapping by defect level mostly decreases with the nanowire diameter, in the range of diameter $d=50- 200$nm of interest. This means, as we increase the nanowire diameter, the average number of electrons trapped by the defect level eventually decreases, which means the system of conduction electrons loses lesser amount of energy so that the electrons effectively stay in the conduction band, remaining energetic there. The hot carrier effect should thus increase with the nanowire diameter for relatively thin nanowires where SRH mechanism dominates over the electron-phonon scattering. This is because the decrease in defect density and defect-induced scattering supersedes the increase in electron-phonon scattering as one increases the diameter of the thin nanowires.
	
	\begin{figure}
		\includegraphics[angle=0,origin=c, scale=0.5]{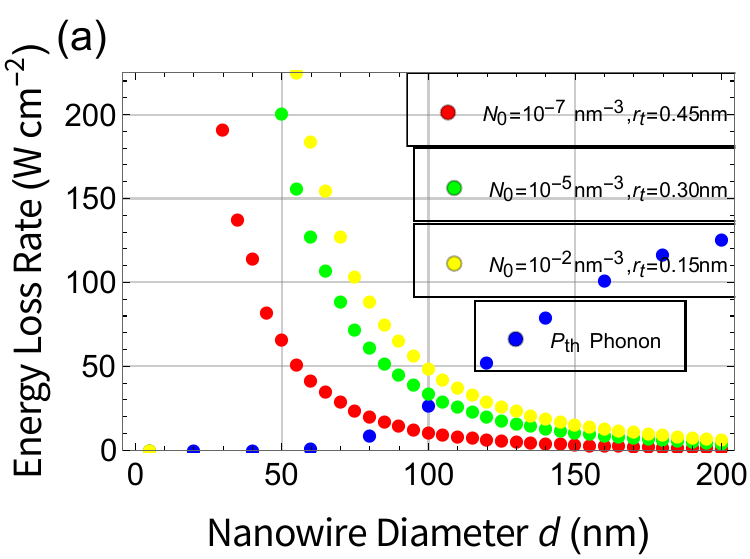}
		\includegraphics[angle=0,origin=c, scale=0.5]{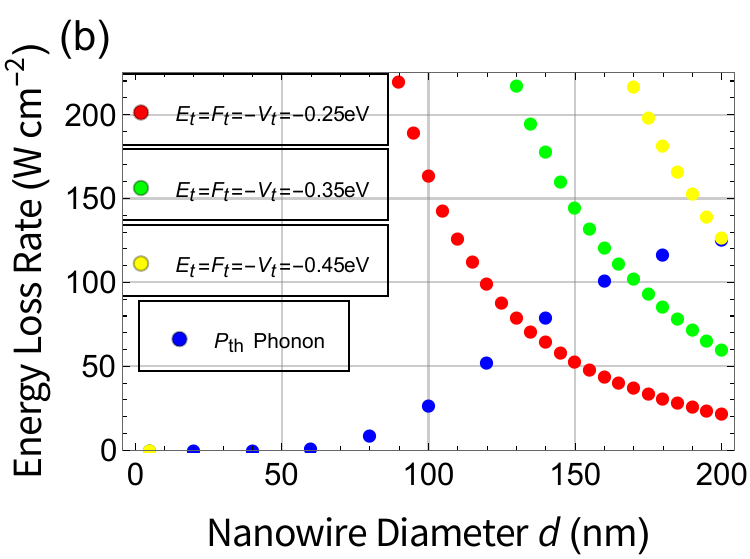}
		\includegraphics[angle=0,origin=c, scale=0.5]{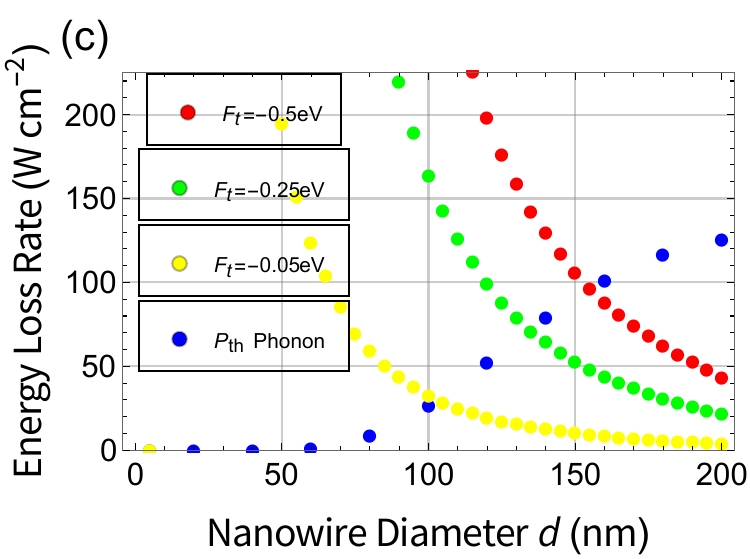}
		\caption{
			The theoretical dependence (decreasing curves)  of the defect-induced net electron energy loss rate $P_{ce}$ on the nanowire diameter of a 1000 nm-long InGaAs/InAlAs core-shell nanowire with different (a)defect densities and associated potential ranges,(b)defect energies, and (c)defect electron Fermi energies. Superposed (blue circles) is the electron-phonon scattering-induced thermalization power $P_{th}$. A relatively wide range of variation for each parameter is considered since its precise numerical value is not precisely known. All other parameters not displayed in the legends are taken from Table I.}\label{fig:Fig3}
	\end{figure}
	
	The crossing point between the thermalization power $P_{th}$ (blue circles) due to electron-phonon scattering within the conduction band and energy loss rate $P_{ce}$ (black circles) due to electron trapping by defect level in Fig. \ref{fig:EnergyLossRates}  corresponds to an ``optimal diameter" where the hot carrier effect is maximized. To see this, the thermalization power or energy loss rate in general can be expressed as \cite{BrisGuillemoles,GiteauJAP}
	\begin{equation}
		P_{th}=Q(T_c-T_L)=\frac{T_c-T_L}{\mathcal{R}}
	\end{equation}
	where $Q$ is the so-called thermalization coefficient while $\mathcal{R}$ is what we call thermalization reluctance. Given a fixed set of values $T_L,T_c$, it is clear that slower rate of thermalization and thus larger reluctance to thermalization should give smaller thermalization power $P_{th}$, eventually leading to larger $\Delta T=T_c-T_L$ if we reverse the role; we fix the $P_{th}$ and maximizes the reluctance $\mathcal{R}$. 
	The crossing point marks the diameter at which the strengths of the two mechanisms; defect-induced scattering of electron and electron-phonon scattering balance each other. Below the optimal diameter, the defect mechanism dominates while above, the electron-phonon scattering takes over. As a result, there is an effective minimum in the electron overall energy loss rate at this optimal diameter, implying a maximum in hot carrier effect. 
	
	The opposite trends of the energy loss rates on the nanowire diameter from the electron-phonon scattering and defect-induced electron scattering predicted from our theory, suggests a strong, non-monotonic dependence of the hot carrier effect on the nanowire diameter. This explains a recent experimental observation \cite{MahyarACSnanoenergymaterials} which indicates the non-monotonic dependence of the hot carrier effect on the cross sectional size of the solar cell nanowire as one increases the diameter in the range between 100 and 200 nm. Equally importantly, our theory predicts a sharp rise to a peak hot carrier effect (in terms of $\Delta T=T_c-T_L$) from below the optimal diameter, due to the sharp drop of $P_{ce}$ in this regime, which agrees as well with the data in \cite{MahyarACSnanoenergymaterials}. 
	
	Of particular interest as well is the dependence of the electron energy loss rate on the parameters of the defect. With details given in the Supplementary Materials \cite{SupplementaryMaterials}, it is found that energy loss rate $P_{ce}$ increases with larger potential range $r_t$ and defect density $N_t$, measured in terms of the product $N_tr^4_t$, as evident from Eqs.(\ref{electroncaptureRateFinal}-\ref{IntegralFactor}) and illustrated in Fig.\ref{fig:Fig3}(a). This dependence covers several orders of magnitude in defect density, implying the robustness of the effect (the balance between the two mechanisms) our theory predicts over wide range of defect density reported in the literature \cite{Defect1, Defect2, Defect3, Defect4}. According to these reports, the typical point defect candidates with such variable density and trap-state energy considered here, are, for example, As vacancies, As antisites, as well as impurity-vacancy complexes. The latter are also considered to be linked to stacking defects present in the nanowires, which are sources for increased impurity defect incorporation \cite{MahyarACSnanoenergymaterials, MahyarPRB2024}. On the other hand, the energy loss rate $P_{ce}$ increases with lower (i.e. deeper) defect level $E_t$ and lower Fermi level for defect electron $F_t$, as illustrated in Fig.\ref{fig:Fig3}(b-c). The first is consistent with general knowledge that a deep level controls the non-radiative life time of charge carriers \cite{Keldysh}, while the second consistent with the fact that lower occupancy of defect level means the latter is capable of accommodating more electrons to be trapped. Further details on the dependence of the electron energy loss rate $P_{ce}$ on the defect parameters is given in \cite{SupplementaryMaterials}. Overall, while the defect parameters are not easy to determine or measure experimentally,  Fig.\ref{fig:Fig3} demonstrates that varying the those parameters over realistic range of intervals or order of magnitude, the optimal diameter remains within the range of 50 - 200 nm, indicating the robustness of mechanism for hot carrier effect optimization.
	
	In summary, we have done a comprehensive study considering various mechanisms involved in the rates of hot carrier thermalization in nanostructures, in nanowires in particular. One is the electron-phonon scattering that leads to electron energy loss increasing with larger nanowire diameter while the other is electron trapping by defect level, resulting in electron energy loss that decreases with larger nanowire diameter. Our results suggest that while smaller nanowire diameter enhances quantum confinement effect, there is an optimal diameter at which hot carrier effect is optimized, in the presence of defects invariably introduced during fabrication process. This conclusion is obtained by extending Shockley-Read-Hall (SRH) original bulk theory to include discrete energy levels of confined electrons. The intrinsic diameter dependence of confined electron density of states in nanowire employed in our theory is crucial in giving the right diameter dependence of electron-defect scattering rate, which cannot be obtained from SRH original theory. Although the focus of this study is on nanowires due to their favorable properties for hot carrier solar cells, our approach can be applied to various nanostructures and provides us with a better understanding of the influence of different mechanisms involved in the rates of hot carrier thermalization in these structures.
	
	\textit{Acknowledgements.\textemdash}H. E. and G. K. acknowledge the funding received from the European Union’s Horizon 2020 research and innovation program under the Marie Skłodowska-Curie Grant Agreement 899987 and the Deutsche Forschungsgemein-schaft (DFG) via Germany’s Excellence Strategy-EXC2089/1-390776260 (e-conversion).

\begin{widetext}
	
	\centering
	\textbf{Supplementary Materials: Interplay of Electron Trapping by Defect Midgap State and Quantum Confinement to Optimize Hot Carrier Effect in a Nanowire Structure}\\
	\centering
	I. Makhfudz$^{1}$, H. Esmaielpour$^{2}$, Y. Hajati$^3$, G. Koblmüller$^{2}$, and Nicolas Cavassilas$^1$
	
		\centering
	$^{1}$IM2NP, UMR CNRS 7334, Aix-Marseille Universit\'{e}, 13013 Marseille, France\\
$^{2}$Walter Schottky Institut, Technische Universität München, Am Coulombwall 4, D-85748 Garching, Germany\\
$^3$Department of Physics, Faculty of Science, Shahid Chamran University of Ahvaz, 6135743135 Ahvaz, Iran

\date{\today}
\bigskip

\begin{flushleft}
	In these Supplementary Materials, we provide the derivation of electron-phonon scattering rate in nanowire geometry and the electron capture probability using partial wave technique. The dependence of defect-induced electron energy loss rate vs diameter on the properties of the point defect (its energy, Fermi level (filling), density, and potential range) which is of utmost interest, is presented at the end.
\end{flushleft}

\end{widetext}

\section{Derivation of Electron-Phonon Scattering Rate in Nanowire}


In kinetic theory, the electron-phonon scattering rate reads 
$ \frac{d N_{{\bf q}_\perp,q _z}}{dt} \Bigr|_{c-LO} = \frac{2 \pi}{\hbar} |M_{{\bf q}_\perp,q _z} |^2 A_{\mathbf{q}_{\perp},q_z}$ where $\hbar$ is the reduced Planck constant and $M_{\mathbf{q}_\perp,q_z}$ the electron-phonon coupling function . The scattering function is given by 
\begin{widetext}
\[
A_{\mathbf{q}_{\perp},q_z}=2\sum_{k_z,k'_z}\sum_{\mathbf{k}_{\perp},\mathbf{k}'_{\perp}}I^2_{\mathbf{k},\mathbf{k}'}((N_{\mathbf{q}_{\perp},q_z}+1)f_{\mathbf{k}_{\perp},k_z}(1-f_{\mathbf{k}'_{\perp},k_z-q_z})
-N_{\mathbf{q}_{\perp},q_z}f_{\mathbf{k}'_{\perp},k_z-q_z}(1-f_{\mathbf{k}_{\perp},k_z}))
\]
\begin{equation}\label{TheSum1}
	\times    \delta(E_{\mathbf{k}_{\perp},k_z}-E_{\mathbf{k}'_{\perp},k_z-q_z}-\hbar\omega_{\mathbf{q}_{\perp},q_z}) \ 
\end{equation}
\end{widetext}
with parameters appearing within the sum are as defined in the main text.  We adopt the notation ${\bf k} = ({\bf k}_\perp, k_z)$ and the same for ${\bf q}$. Henceforth we suppose that there is no dispersion in the LO phonon spectrum, thus $\omega_{\mathbf{q}_{\perp},q_z} =\omega_{LO}$. Within effective mass approximation
\begin{equation}\label{ChangeinKineticEnergy}
E_{\mathbf{k}_{\perp},k_z}=\frac{\hbar^2(\mathbf{k}^2_{\perp}+k^2_z)}{2m_c}.
\end{equation}
The change in the electron kinetic energy between before and after the scattering by the phonon then reads
\begin{equation}
E_{\mathbf{k}_{\perp},k_z}-E_{\mathbf{k}'_{\perp},k_z-q_z}    =\frac{\hbar^2}{2m_c}\left[(k^2_{\perp}-{k'_{\perp}}^2)-q^2_{z}+2k_{z}q_{z}
\right] \ .
\end{equation}
Since $I^2_{\mathbf{k},\mathbf{k}'} \sim  \delta(k'_z-k_z+q_z)$, the sum $ \sum_{k'_z}$ immediately imposes $k'_z=k_z-q_z$. On the other hand, for the remaining  sum $ \sum_{k_z}$, we use 
\begin{equation}\label{transformationsumtointegral}
\sum_{k_z} ...\rightarrow \frac{L}{2\pi}\int^{\infty}_{-\infty} d k_z \ ... \ ,
\end{equation}
where $L$ is the length of the nanowire. Expressing the discrete nature of the confined electron transverse wave vector $k_{\perp}(\mathbf{n})=\alpha_{ln}/R,k'_{\perp}(\mathbf{n}')=\alpha_{l'n'}/R$ where $\mathbf{n}=(n,l), \mathbf{n}'=(n',l')$, $\alpha_{l,n}$ is the $n^{th}$ zero of the Bessel function of the first kind of the $l^{th}$ order $J_l(\cdots)$; i.e. $J_l(\alpha_{l,n})=0$, while leaving the LO phonon longitudinal wave vector $q_z$ as it is, we get
\begin{widetext}
\[
A_{\mathbf{q}_{\perp},q_z}=\frac{2 m_c}{\hbar^2 q_z} \sum_{\mathbf{n},\mathbf{n}'}\frac{L}{2\pi}\int^{\infty}_{-\infty} d k_z
\left[(N_{\mathbf{q}_{\perp},q_z}+1)f_{\mathbf{k}_{\perp}(\mathbf{n}),k_z}(1-f_{\mathbf{k}'_{\perp}(\mathbf{n}'),k_z-q_z})
-N_{\mathbf{q}_{\perp},q_z}f_{\mathbf{k}'_{\perp}(\mathbf{n}'),k_z-q_z}(1-f_{\mathbf{k}_{\perp}(\mathbf{n}),k_z})\right]
\]
\begin{equation}\label{rateofchangeofbosonnumber1}
	\times I^2_{\mathbf{k}_{\perp}(\mathbf{n}),k_z;\mathbf{k}'_{\perp}(\mathbf{n}'),k_z-q_z} \delta\left(k_z - \frac{q_z}{2}+  \frac{1}{2q_zR^2}\left(\alpha^2_{ln}-\alpha^2_{l'n'}\right)
	- \frac{m_c \omega_{LO} }{\hbar q_z}
	\right)
\end{equation}
\end{widetext}
involving a Dirac delta function of $k_z$.

The terms in the square bracket in Eq.~(\ref{rateofchangeofbosonnumber1}) can be simplified using an identity \cite{Wurfel},
\begin{equation}
f_{\mathbf{k}_{\perp}(n),k_z}(1-f_{\mathbf{k}'_{\perp}(n'),k'_z}) =N_{\mathbf{q}}(T_c)(f_{\mathbf{k}'_{\perp}(n'),k'_z}-f_{\mathbf{k}_{\perp}(n),k_z}) \ ,
\end{equation}
where $N_{\mathbf{q}}(T_c)=1/(\exp(\hbar \omega_{LO}/(k_BT_c))-1)$ is the equilibrium phonon distribution at the carrier temperature $T_c$.
Gathering the previous results, we finally get
\begin{widetext}
\begin{equation}\label{aqqz}
	A_{\mathbf{q}_{\perp},q_z}
	=( N_{\mathbf{q}}(T_c) - N_{\mathbf{q}_{\perp},q_z}) \frac{m_c L}{\pi\hbar^2 q_z} \sum_{\mathbf{n},\mathbf{n}'}\int^{\infty}_{-\infty} d k_z I^2_{\mathbf{k}_{\perp}(\mathbf{n}),k_z;\mathbf{k}'_{\perp}(\mathbf{n}'),k_z-q_z} 
	(f_{\mathbf{k}'_{\perp}(n'),k'_z}-f_{\mathbf{k}_{\perp}(n),k_z}) \delta(k_z-k^0_z) \, ,
\end{equation}
\end{widetext}
where
\[
k^0_z=\frac{1}{2q_z}\left(\mathbf{k}'^2_{\perp}+q^2_z-\mathbf{k}^2_{\perp}\right)+\frac{m_c}{\hbar^2q_z}\hbar\omega_{LO}
\]
\begin{equation}
= \frac{q_z}{2}-  \frac{1}{2q_zR^2}\left(\alpha^2_{ln}-\alpha^2_{l'n'}\right)
+ \frac{m_c \omega_{LO} }{\hbar q_z}.
\end{equation}

Finally, defining the scattering time $\tau^{c-LO}_{\bf{q}}$ by
\begin{equation}
\frac{2 \pi}{\hbar} |M_{{\bf q}_\perp,q _z} |^2 A_{\mathbf{q}_{\perp},q_z} = \frac{(N_{\bf{q}}(T_c) -N_{\bf{q}})}{\tau^{c-LO}_{\bf{q}}}
\end{equation}
and performing the integration over $k_z$ in Eq.(\ref{aqqz}) gives 

\[
\frac{1}{\tau^{c-LO}_{\mathbf{q}}}=
\]
\begin{equation}
\frac{2Lm_c}{\hbar^3q_z}|M_{\bf{q}_{\perp},q_z}|^2\sum_{\mathbf{k}_{\perp},\mathbf{k}'_{\perp}}I^2_{\mathbf{k}_{\perp},k^0_z;\mathbf{k}'_{\perp},k^0_z-q_z}\left(f_{\mathbf{k}'_{\perp},k^0_z-q_z}-f_{\mathbf{k}_{\perp},k^0_z}\right)
\end{equation}
as displayed in the main text.


\section{Derivation of Electron Capture Probability}
Detailed balance analysis does not provide an explicit expression for the probability function $c_l(E)$, which requires an analytical derivation from an appropriate microscopic model. 
To this end, we use the simplest model of scattering theory for the point defect; scattering of an electron by a potential well of depth $V_t$ of finite range $r_t$. We use partial wave technique \cite{LandauLifshitzQMs} to compute the scattering cross section appearing in the SRH expression \cite{ShockleyReads} for electron capture probability
\begin{equation}\label{electroncaptureprobabilityS}
c_l(E)=\langle v(E) A_l(E)\rangle.
\end{equation}
In this case, the electron is trapped by a spherical attractive potential whose length scale is much smaller than nanowire diameter, that is $r_t\ll R$, permitting a simplified picture in terms of plane wave scattering in the bulk analyzed within spherical coordinates. Within this approximation, $A_l$(and hence $c_l$) is independent of $l$ and this subscript is dropped from now on.

The electron capture probability  as originally defined in the SRH theory in Eq.(\ref{electroncaptureprobabilityS}) is now given within the scattering theory by
\begin{equation} \label{electroncaptureprobabilityintegral}
c(E)=\langle v(E) A(E)\rangle\equiv\int d\Omega v(E) \frac{d\sigma(E)}{d\Omega}
\end{equation}
where $d\sigma(E)/d\Omega$ is the differential cross section of the electron while $d\Omega=\sin\theta d\theta d\phi$ in spherical coordinates and
\begin{equation}\label{ElectronVelocity}
v(E)= \sqrt{ \frac{2E}{m_c}}
\end{equation}
within effective mass approximation. In our formulation, the angular integral in Eq.(\ref{electroncaptureprobabilityintegral}) not only gives rise to total scattering cross section $A(E)$ but also an averaging $\langle \cdots \rangle$ of the $v(E) A(E)$ over all directions of scattered electron. 

We follow standard theory of scattering in quantum mechanics based on partial wave approach \cite{LandauLifshitzQMs}, where the electron wave function upon scattering by a potential is given by
\begin{equation}
\psi(r,\theta,\phi)=e^{ikr\cos\theta}+f(\theta,\phi)\frac{e^{ikr}}{r}
\end{equation}
where we have taken an incoming plane wave while  $f(\theta,\phi)$ is the angular function of the outgoing wave function of the electron upon scattering. The formulation of this scattering wave function done in spherical coordinates in terms of plane wave, instead of in cylindrical coordinates in terms of plane wave  
\begin{equation} \label{planewavefunctioninnanowireSM}
\psi(r,\phi,z)\equiv \phi_{\mathbf{n}}(\mathbf{r})=e^{ik_rr}e^{ik_{\phi}r_{\phi}}e^{ik_zz},
\end{equation}
or mathematically precise in terms of the eigenfunctions  \begin{equation} \label{electroneigenfunctioninnanowireSM}
\psi(r,\phi,z)\equiv \phi_{\mathbf{n}}(\mathbf{r})=J_l\left(\frac{\alpha_{l,n}r}{R}\right)e^{il\phi}e^{ik_zz},
\end{equation}
appropriate for cylindical geometry of the nanowire, is an approximation; it is justified assuming defect size and electron de Broglie wave length much smaller than the nanowire radius, making the nanowire effectively looks like a bulk system from the perspective of the electron and point defect. More accurate derivation in terms of the eigenfunctions Eq.(\ref{electroneigenfunctioninnanowireSM})
is not going to change the results qualitatively when the nanowire radius is much larger than the size of the point defect; $R\gg r_t$. 

\begin{figure}
\includegraphics[angle=0,origin=c, scale=0.35]{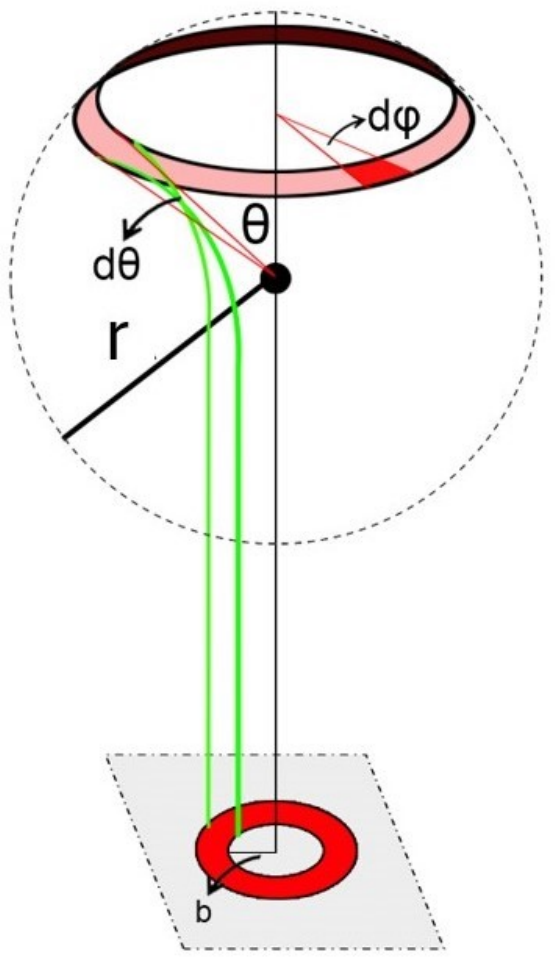}
\caption{
	The geometry of the electron scattering by a spherical defect potential.}\label{fig:FigSM}
\end{figure}

With the geometry of the scattering as illustrated in Fig.\ref{fig:FigSM}, the angular function $f(\theta,\phi)$ determines the differential scattering cross section,

\begin{equation} \label{differentialcrossection}
\frac{d\sigma}{d\Omega}=|f(\theta,\phi)|^2
\end{equation}
at given angles $\theta,\phi$. Considering axial symmetric scattering, the angular dependence on $\phi$ drops out. Within Born approximation, the angular function becomes
\begin{equation} \label{AngularFunctionBornApprox}
f(\theta)=\frac{1}{k}\sum_{l}(2l+1)e^{i\delta_l(k)}\sin\delta_l(k) P_l(\cos\theta) 
\end{equation}
where $\delta_l(k)$ is the phase shift associated with $P_l(\cdots)$ the Legendre polynomial of order $l=0,1,2,\cdots$. For small phase shifts for which the Born approximation is valid, the phase shifts are given by
\begin{equation}\label{SmallPhaseShifts0}
\delta_l(k)\approx -\frac{2m_ck}{\hbar^2}\int^\infty_0 dr r^2V(r)\left(j_l(kr)\right)^2
\end{equation}
where $j_l(\cdots)$ is spherical Bessel function of order $l$.

Now, we assume a microscopic (atomic scale) spherical potential well of radius $r_t$ of depth $V_t$. That is, $V(r)=-V_t$ for $r<R$ and zero otherwise. Intuitively speaking, the $V_t$ can be taken to be the energy difference between the bottom of conduction band $\varepsilon_c$, eventually taken to be the reference zero energy in our calculation, and the defect level. As such, the trapping of the electrons by the defect level means the electrons fall into the bottom of this potential well by losing energy via emission of phonons, a process otherwise outlawed in the absence of the defect level because the final state would be within the forbidden gap. In this case, we only account for the energy loss without further detail on the phonon emission. 

Substituting the $V(r)$ into Eq.(\ref{SmallPhaseShifts0}), one obtains
\begin{equation} \label{SmallPhaseShifts}
\delta_l(k)= \frac{m_cV_t\pi^2r^2_t}{2\hbar^2}\left(J^2_{l+\frac{1}{2}}(kr_t)-J_{l-\frac{1}{2}}(kr_t)J_{l+\frac{3}{2}}(kr_t)\right)
\end{equation}
valid for $l>-3/2$ otherwise the integral does not converge. The $J_{m}(kr_t)$ is Bessel function of the first kind of order $m$. We note that small phase shifts imply shallow potential well with short range. Substituting this result for phase shift into the angular function in Eq.(\ref{AngularFunctionBornApprox}) and making use of small phase shift approximation $\sin\delta_l(k)\approx \delta_l(k)$, we obtain 
\begin{widetext} \begin{equation}\label{ModulusSquaredAngularFunction}
	|f_k(\theta)|^2=\left(\frac{m_cV_t\pi^2}{2p^2_tk}\right)^2\sum_{l,l'}(2l+1)(2l'+1) P_l(\cos\theta)P_{l'}(\cos\theta)\left(J^2_{l+\frac{1}{2}}(\xi_k)-J_{l-\frac{1}{2}}(\xi_k)J_{l+\frac{3}{2}}(\xi_k)\right)\left(J^2_{l'+\frac{1}{2}}(\xi_k)-J_{l'-\frac{1}{2}}(\xi_k)J_{l'+\frac{3}{2}}(\xi_k)\right) 
\end{equation} 
\end{widetext}
where we have exhibited explicitly the $k$-dependence (implying also energy dependence) of the angular function, while defining $p_t=\hbar/r_t$ and $\xi_k=kr_t$. In other words, both sides of Eq.( \ref{differentialcrossection}) depend implicitly on energy $E$. Substituting the expression above into Eq.(\ref{differentialcrossection}) and using the identity
\begin{equation}
\int^1_{-1}  P_l(x)P_{l'}(x)dx=\frac{2}{2l+1}\delta_{ll'}
\end{equation}
leads to the analytical result for $c_n(E)$ presented in the main text.
Physical consideration suggests that the phase shift from Eq.(\ref{SmallPhaseShifts0}) be finite. 

While our theory is more appropriate for deep level, it may also be applied to approximately describe shallow level, for which case $F_t$ is just few tens of meV under $F_n$ and $r_t$ would be much larger, reflective of Coulombic nature of impurity potential that normally gives shallow level. In our formulation, we have approximated the potential of the point defect, which can be Coulombic in reality going as $1/r$ for shallow level and is thus singular at $r=0$, with a simple attractive spherical potential well of constant depth $V_t$. Other common potentials such as Lennard-Jones potential also have singularity at zero radius. It is therefore imperative to remove this singularity, as it gives rise to diverging phase shift, which is non physical. This physical consideration imposes that the Bessel function entering the integral giving the phase shift in Eq.(\ref{SmallPhaseShifts0}) be zero at $r=0$. This is satisfied only for $l,l'>0$ among $l,l'=0,1,2,\cdots$. The sums over $l,l'$ in Eq.(\ref{ModulusSquaredAngularFunction}) thus go from 1 to $\infty$.

\begin{figure*}
\centering
\begin{minipage}[b]{.4\textwidth}
	\includegraphics[angle=0,origin=c, scale=0.5]{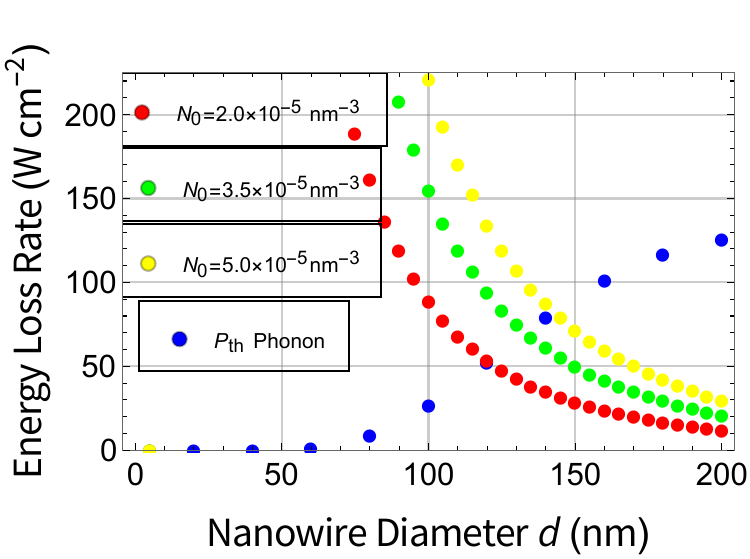}
	\caption{
		The theoretical dependence (decreasing curves)  of the defect-induced net electron energy loss rate $P_{ce}$ on the nanowire diameter of a 1000 nm-long InGaAs/InAlAs core-shell nanowire with diameter-dependent defect density ansatz, for different defect densities $N_0=2.0\times 10^{-5}$/nm$^3$(red), $N_0=3.5\times 10^{-5}$/nm$^3$(green), $N_0=5.0\times 10^{-5}$/nm$^3$(yellow) respectively. All the other parameters are given in Table I of the main text. Superposed (blue circles) is the electron-phonon scattering-induced thermalization power $P_{th}$.}\label{fig:EnergyLossRatesDefectDensity}
\end{minipage}\qquad
\begin{minipage}[b]{.4\textwidth}
	\includegraphics[angle=0,origin=c, scale=0.5]{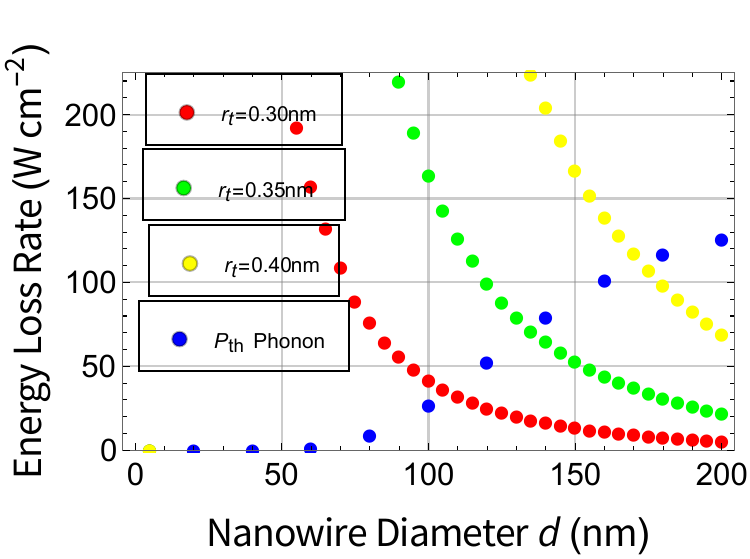}
	\caption{
		The theoretical dependence (decreasing curves)  of the defect-induced net electron energy loss rate $P_{ce}$ on the nanowire diameter of a 1000 nm-long InGaAs/InAlAs core-shell nanowire with  different defect potential ranges $r_t=0.08$nm(red), $r_t=0.10$nm(green), $r_t=0.12$nm(yellow) respectively. Superposed (blue circles) is the electron-phonon scattering-induced thermalization power $P_{th}$.}\label{fig:EnergyLossRatevsImpurityDefectPotentialRange}
\end{minipage}
\end{figure*}

\section{Dependence of Energy Loss Rate on Defect Density, Energy, and Potential Range}

In this section, we present the dependencies of the defect-induced electron energy loss on the defect parameters, which are the principal interest in this work.

First, we vary the point defect density. Fig.\ref{fig:EnergyLossRatesDefectDensity} shows that increasing defect density increases the electron energy loss rate. This is very intuitive and comes simply from the fact that the energy loss rate is proportional to the defect density.

Secondly, we vary the defect energy parameters; the energy of the defect level $E_t$, its Fermi energy $F_t$, and the spherical potential well depth $V_t=-E_t$. First, we fix the Fermi energy to be always equal to the defect level energy $F_t=E_t$. The result is presented in Fig. 4 in the main text. The result indicates that lowering the defect level energy and its Fermi energy at the same time tends to increase the electron energy loss. At first it may seem that lowering the defect level energy will make it less likely to capture electrons from the conduction band. But it should be noted that lower defect level energy also reduces the probability of electron re-emission back into the conduction band. This is because, once an electron is lost from being captured by the defect level, with lower defect level energy, the more likely the electron stays stays there than to be re-emitted back to the conduction band. In addition, the lower the defect level energy, the larger the energy loss $E-E_t$ is for each electron capture process, thus raising the energy loss rate. Our result is also consistent with general understanding that shallow defect level, that is, one that is not too far from the bottom of the conduction band, does not give rise to strong energy loss.

Next, we fix the defect energy level, but vary its occupation by varying the defect Fermi energy level. The result is presented in Fig. 4 in the main text. The electron energy loss rate in fact decreases with the defect Fermi energy level. This is clearly because the higher the defect Fermi energy level, the higher the occupation of the defect level, which means the lesser the state space available for the defect level to accommodate new electrons to be captured. The dependence on these energy parameters is rather strong due to the fact that they enter exponential function via Fermi-Dirac distribution function.

Finally, we present the dependence of the electron energy loss rate on the range $r_t$ of the defect potential. According to Eq.(22) in the main text, the leading order dependence on $r_t$ is $P_{ce}\sim r^4_t$, which means the electron energy loss rate rapidly increases with the increases in the defect potential range. This is very intuitive because longer range potential should be able to capture more electrons. However, there is also hidden dependence on $r_t$ via the function $G_{l'}(E)$ in Eq.(19) in the main text. The full dependence is displayed in Fig.\ref{fig:EnergyLossRatevsImpurityDefectPotentialRange}which shows the dominant $r^4_0$ dependence on the i,purity potential range; the electron energy loss rate strongly increases with the increase in defect potential range.


\begin{thebibliography}{1}
\bibitem{NoteDefect}Our theory on defect scattering of electron is general enough to be applicable to any type of point or point-like defect whereever it occurs in the nanowire, which may be present due to unintentional doping \cite{dopingpointdefect} or degradation in crystalline integrity \cite{degradationcrystal}, as long as it leads to defect level that can trap electrons \cite{trap1}\cite{trap2} as necessary for the SRH theory \cite{ShockleyRead} to apply.
\bibitem{NoteDefectDensity}The $N_t$ cannot be taken as constant because in practice, defect density strongly depends on the size of the nanostructure; increasing with the decrease in nanowire diameter \cite{FacileACSNano, RuhstorferJAP}. A simple model with diameter-dependent defect density is assumed such that the number of defects per unit length $N^{def}/L$ is constant, so that only diameter dependence remains, leading to diameter-dependent $N_t$ that satisfies $d^2N_t(d)=d^2_0N_0=4N^{def}_0/(\pi L_0)$, where $d_0, N_0$ are reference diameter and defect densities respectively, while $N^{def}_0$ and  $L_0$ are the corresponding reference  total number of defects and nanowire length, defined at the largest diameter (200nm in our study),  from which defect density $N_t$ at smaller diameters $d$ is determined. More realistic but delicate situation where $N^{def}/L$ varies (decreases) with diameter does not change our final results qualitatively.
\bibitem{NoteParameters}Some of these parameters such as conduction electron Fermi level $F_n$ may themselves depend on diameter (e.g. if the electron density is fixed), but the variation is small enough to not give dramatic changes \cite{MakhfudzJoPD}.
\end{thebibliography}
\end{document}